# Granular Ta-Te nanowire superconductivity violating the Pauli limit

Lingxiao Zhao[1#], Yi Zhao[1#], Cuiying Pei[1], Changhua Li[1], Qi Wang[1,2], Juefei Wu[1], Weizheng Cao[1], Lin Xiong[1,3], Haiyin Zhu[1,3], Tianping Ying[4], Yanpeng Qi[1,2,3]*

1. School of Physical Science and Technology, ShanghaiTech University, 201210 Shanghai, China.

2. ShanghaiTech Laboratory for Topological Physics, ShanghaiTech University, 201210 Shanghai, China

3. Shanghai Key Laboratory of High-resolution Electron Microscopy, ShanghaiTech University, 201210 Shanghai, China

4. Institute of Physics and University of Chinese Academy of Sciences, Chinese Academy of Sciences, 100190 Beijing, China

\# These authors contributed to this work equally.

\* Correspondence should be addressed to Y.Q. (qiyp@shanghaitech.edu.cn)

**Abstract:** Strategies to achieve higher upper-critical-field superconductors ($\mu_0H_{c2}(0)$) are of great interest for both fundamental science and practical applications. While reducing the thickness of two-dimensional (2D) materials to a few layers significantly enhances $\mu_0H_{c2}(0)$ with accompanied potential unconventional pairing mechanisms, further dimensional reduction to 1D compounds rarely exceeds the expected Pauli limit. Here, we report the discovery of a 1D granular Ta-Te nanowire that becomes superconducting under high pressure, with a maximum critical temperature ($T_c$) of 5.1 K. Remarkably, the $\mu_0H_{c2}(0)$ reaches 16 T, which is twice the Pauli limit, setting a record of $\mu_0H_{c2}(0)$ in all the reported 1D superconductors. Our work demonstrates that the Ta-Te nanowire not only is a potential candidate for applications in high magnetic fields, but also provides an ideal platform for further investigations of the mechanisms between nanowires and large $\mu_0H_{c2}(0)$.

## Introduction

Since the discovery of the Meissner effect, efforts for pursuing superconductors with higher upper critical fields ($\mu_0H_{c2}(0)$) have never been paused. In recent years, the success in mechanical exfoliation of 2D van der Waals materials has unveiled many new superconductors with large upper critical fields, especially in nano-films of transition metal chalcogenides (TMCs). Few-layer NbSe$_2$ is reported to be an Ising superconductor and exhibits orbital Fulde-Ferrell-Larkin-Ovchinnikov (FFLO) states under large magnetic fields[1-5]. As a result, the in-plane $\mu_0H_{c2}(0)$ of few-layer NbSe$_2$ is about 1.5 times of Pauli limit ($\mu_0H_p$). Similar behaviors are also observed in MoS$_2$[6], TaS$_2$[7] and gated WS$_2$[8].

The discovery of large upper critical fields in 2D materials inspires the exploration of further reducing their dimensions, which may lead to higher $\mu_0H_p$ and possibly unconventional superconductivity. However, reports of superconductivity in 1D nanowires violating the Pauli limit are extremely rare. To the best of our knowledge, Bi nanowires are reported to show complex superconducting properties, depending on the morphology of the nanowires[9-11]. Single crystal Bi nanowires' $T_c$ and $\mu_0H_{c2}(0)$ are 1.3 K and 4 T, respectively[10]. In contrast, granular nanowires of Bi exhibit 8.3 K and 4.5 T, which show similar behavior to the bulk high-pressure phase, Bi-V[9]. Another example is the B nanowire, which is insulating at ambient pressure, but superconducting at 84 GPa. The $\mu_0H_{c2}(0)$ of B nanowires is about 2.5 T while the $T_c$ is only about 1.5 K[12]. In the cases of single crystal Bi nanowires and pressurized B nanowires, the upper critical fields are quite large considering their low $T_c$. Apart from high pressure, proximity effects and ion beams can also induce superconductivity in nanowires[13-15]. Therefore, the dependences of the critical behavior of superconducting nanowires on morphology, pressure, gating, proximity effect, and synthesis methods are quite complex, little consensus has been reached up to now.

In this work, we synthesized a novel TMC nanowire composed of Ta$_3$Te$_4$ granules. The nanowires show semi-metal behavior at ambient pressure and superconductivity under high pressure. The upper critical field reaches about 16 T at 21 GPa, which is about 4 times $T_c$ and much larger than the value of Pauli limit. The $\mu_0H_{c2}(0)$ - $T_c$ ratio is the highest among all the reported superconducting nanowires. We have successfully combined the advantages of TMC compounds, nanowires and high-pressure modulations, realizing a large upper critical field in compressed Ta-Te nanowires.

Ta-Te nanowires are synthesized by the chemical vapor transport (CVT) method[16]. A Stoichiometric amount of 99.999% tantalum and tellurium powder (Alfar Aesar) is mixed and sealed in evacuated silica ampoules. The starting powder in each ampoule weighs 1 gram and 20 mg of iodine as the transport agent. The size of the ampoules used is 10 cm (length) * 1.5 cm (diameter). The sealed ampoules are heated to 1100 °C in 10 hours, and annealed for 24 hours in a box furnace. The small temperature gradient in the box furnace is enough for the growth of the nanowires. The nanowires are black, and show fiber-like texture, with typical dimensions of 100 nm * 100 nm * 100 μm. They are densely tangled, forming bundles, as shown in Figure S1.

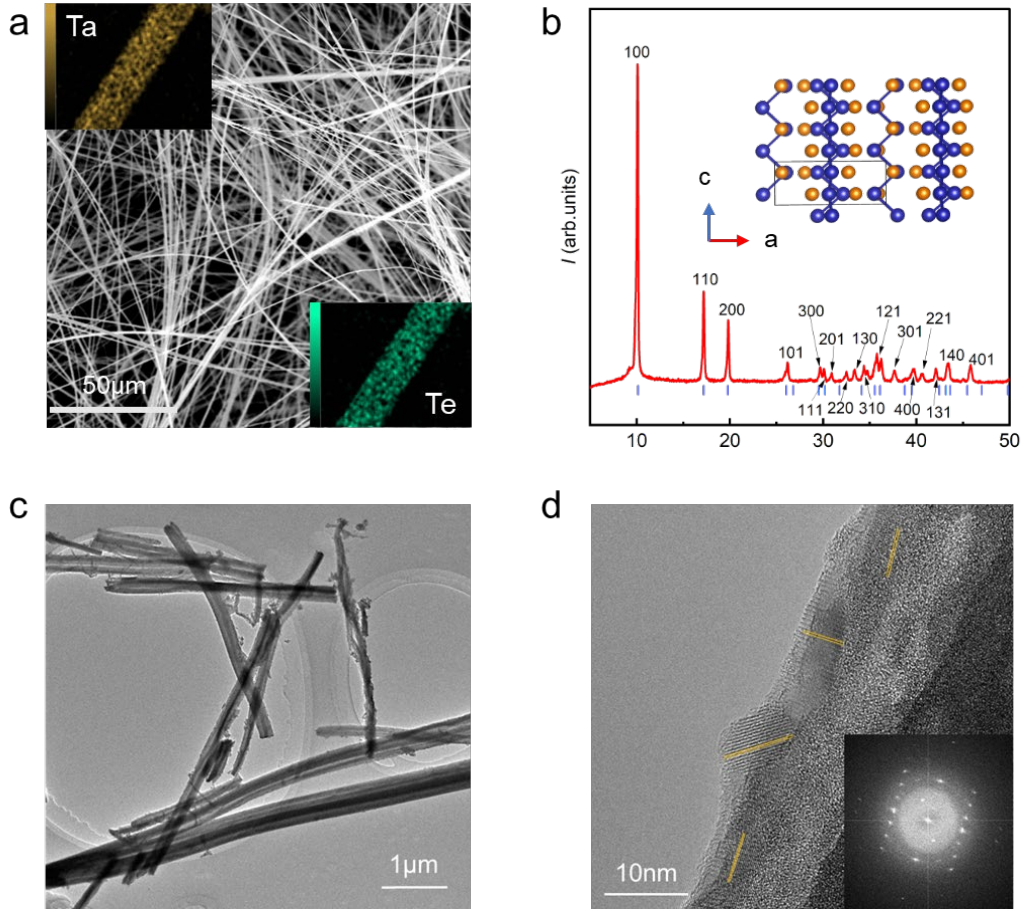

Figure 1. Structural characterization of Ta-Te nanowires at ambient pressure. (a) The SEM image of the nanowires under low magnification. The insets show elemental mapping under SEM. (b) Powder XRD results of a bunch of nanowires. The inset shows the crystal structure of Ta-Te nanowires. (c) The TEM image of the nanowires under low magnification. (d) The TEM images of a nanowire under high magnification. The inset shows the fast Fourier transform (FFT) of the images.

A low magnification SEM image of the synthesized samples is presented in Figure 1a. Under higher magnification, we performed elemental dispersion spectrum (EDS) mapping of a single nanowire. The results are exhibited in the insets of Figure 1a. We observe that Ta and Te atoms are uniformly distributed in the nanowire. We have also conducted powder X-ray diffraction (XRD) measurements on a bunch of nanowires directly acquired from synthesis, as shown in Figure 1b. Though the strongly anisotropic orientation of the nanowires causes difficulty in Rietveld refinements, we have confirmed that the indexed XRD peaks are correlated to the hexagonal $Nb_3Te_4$ structure[17], with the space group $P6_3/m$, as shown in the inset of Figure 1b. It is worth noting that the average full width at half maximum of the diffraction peaks is 0.2°. Using the Scherrer equation, the estimated grain size is 58 nm, implying that these 1D samples may be composed of numerous small nanocrystals.

To further investigate the appearance and morphology of the nanowires, we have performed high magnification SEM and TEM observations. From Figure 1a, we can observe that the nanowires directly collected from synthesis are densely tangled. The samples we used for the following high-pressure *in-situ* resistivity measurements are of the same appearance as the bundle of nanowires near the right edge of Figure S1. The nearly isolated nanowire in the middle of Figure S1 is used for resistivity measurements at ambient pressure.

The samples for TEM are prepared by dispersion of nanowires into ethanol using ultrasonic methods. The dispersed nanowires are transferred to copper grids with carbon membranes. As shown in the low magnification image in Figure 1d, the dispersed samples stand on the carbon membranes. Some of the nanowires are slightly distorted or twisted due to local stress.

High-resolution TEM image of a single nanowire is shown in Figure 1d. The nanowires are composed of crystalline granules. The orientations of the crystalline granules are slightly random, as indicated by yellow lines in Figure 1d. The insets show the fast fourier transformations (FFT) of the images. In the FFT results, "amorphous" halos are observed. This indicates that the grain boundaries exhibit disordered structure. The nanowires thus have bamboo-like structures.

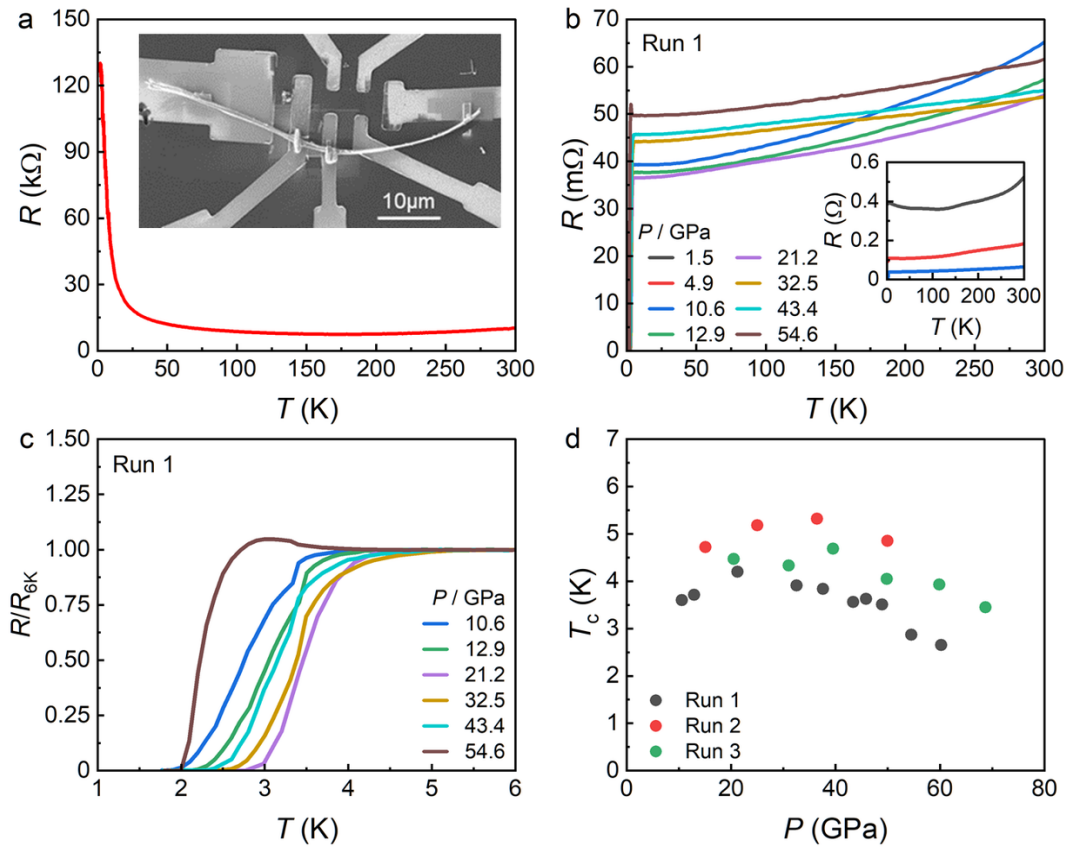

Figure 2. Resistivity measurements of Ta-Te nanowires. (a) The temperature dependence of resistivity at ambient pressure. The inset shows the setup of electrodes

and focused ion beam (FIB) transported sample. (b) The temperature dependence of resistivity under 1.5-54.6 GPa. (c) The normalized temperature dependence of resistivity at low temperature. (d) The $T_c$-Pressure phase diagram.

The temperature dependence of resistivity of a single nanowire at ambient pressure is summarized in Figure 2a. The sample is transported to pre-set electrodes by focused ion beam (FIB) under SEM (Inset of Figure 1e). We use ionic deposited Pt to contact the sample with the electrodes. The nanowire shows an insulator-type electronic transport property, as shown in Figure 2a. The resistivity first slightly drops with cooling temperature. At about 200 K, it turns to increase rapidly, showing insulating behavior. This result is quite different from the metallic behavior of its isomer $Nb_3Te_4$[18].

The electronic structure of the sample is quite important, given that the nanowires show semi-metal type transport properties. We have performed IR transmission spectrum on Ta-Te nanowires to deduce the band gap, shown in Figure S2. The transmission rate of IR light with wavelengths from 2 to 15 um is lower than 1%. So, the optical band gap is at least lower than 75 meV. We suggest that the insulating behavior of the nanowire below 200 K is not a result of the electronic band structure.

Since the samples are insulating at ambient pressure, we attempted utilizing high-pressure to induce superconductivity in Ta-Te nanowires. We have measured the temperature dependence of resistivity from 1.8 K to 300 K under high pressure, as shown in Figure 2b-c. Bunches of tangled nanowires are directly loaded into the sample chamber. At 1.5 GPa, the sample shows insulating behavior, similar to ambient state. It transforms into bad-metal type at 4.9 GPa. During compression, the resistivity at room temperature gradually decreases. At 10.6 GPa, superconductivity onsets at about 3.5 K. The delta $T_c$ is rather wide at this pressure point, with the resistivity dropping to zero at 2 K. With further compression, $T_c$ shows a dome-like evolution with a maximum of 4.13 K at 21.2 GPa, and it drops slowly to about 2.7 K at the highest-pressure point measured in this work. Other runs of measurements have acquired similar evolutions, as shown in Figure S3-4. The phase diagram of $T_c$ and pressure from multiple runs are summarized in Figure 2d.

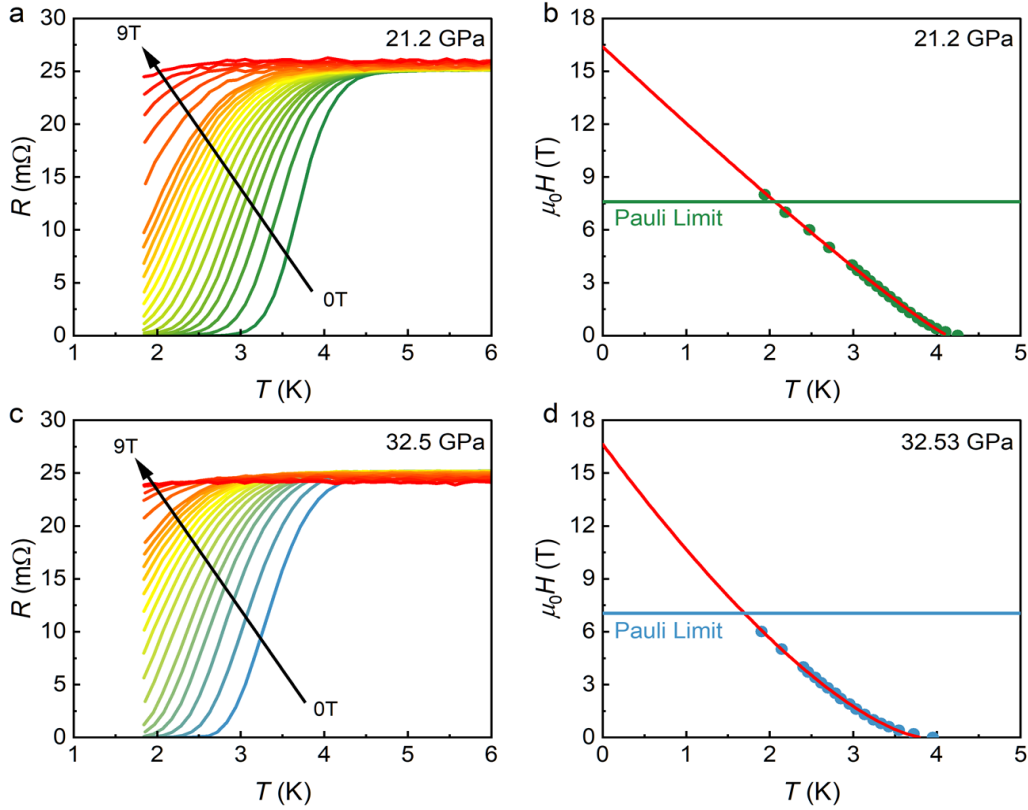

Figure 3. The upper critical field of Ta-Te nanowires. (a) The temperature dependence of resistivity under different magnetic fields from 0 T to 9 T, at 21.2 GPa. (b) The fitted curve of $\mu_0H_{c2}(T)$ with multi-band model at 21.2 GPa. (c) and (d) are similar to (a) and (b), but at 32.5 GPa. The green line in (b) and blue line in (d) denotes the Pauli limit.

To further investigate the superconductivity properties of Ta-Te nanowires, we have measured the upper critical field of the sample at 21.2 GPa and 32.5 GPa (Figure 3). In Figure 3a and c, as the magnetic field increases, $T_c$ gradually decreases until it disappears above the lowest measured temperature. By defining the temperature where the resistance drops by 90% under the magnetic field as $T_c$, we plot the $\mu_0H_{c2}$ against $T_c$ in Figure 3b and d. We then fit the data using the multi-band model. We have found that at 21.2 GPa and 32.53 GPa, the $\mu_0H_{c2}(0)$ are 16.39 T and 16.64 T, respectively. It should be noted that the Pauli limit $\mu_0H_p$ at these two pressure points are 7.60 T and 7.05 T, respectively. The $\mu_0H_{c2}(0)$ results are much higher than $\mu_0H_p$, exhibiting exotic superconducting behavior.

The Pauli limit violating behavior indicates that the pressure-induced superconductivity in Ta-Te nanowires possibly originates from a complex mechanism. According to literature, the $\mu_0H_{c2}(0)$ of most superconducting nanowires are near or below the Pauli limit, regardless of the $T_c$ (Figure 4). Compared with them, Ta-Te nanowires show significantly larger $\mu_0H_{c2}(0)$. In the case of Ta-Te nanowires, the $\mu_0H_{c2}(0)$ is larger than two times $\mu_0H_p$. Some 2D TMCs present similar phenomena. For example, few-layer NbSe$_2$ is an Ising superconductor, with large in-plane $\mu_0H_{c2}(0)$. It also turns into FFLO state at low temperature and high magnetic field[5], with an abrupt upturn of $\mu_0H_{c2}$ after the transition point in the phase diagram. The large upper critical field of the Ta-Te nanowires may originate from a similar mechanism, since the compositional element Ta is moderately heavy, with strong spin-orbital coupling. However, no signals of upturning FFLO states are observed under 9 T in this work (Figure 3b and d). Large upper critical fields have also been observed in some other unconventional superconductors such as cuprates, iron arsenides, heavy-fermion superconductors, etc[19-29]. The exotic pairing mechanisms play important roles in enhancing $\mu_0H_{c2}(0)$. In short, further investigations on the Ta-Te nanowires are required to unveil the pairing mechanism and its influence on the large $\mu_0H_{c2}(0)$. Apart from the pairing mechanisms of large $\mu_0H_{c2}(0)$ observed in these materials, there might be exotic unknown origin in the 1D nanowires. In addition, due to the large upper critical field (16 T) and the moderate $T_c$ (4 K) among known superconducting nanowires, application potentials of Ta-Te nanowires can be expected, considering that the nanowires are easy to synthesize, and may exhibit excellent mechanical properties.

In summary, we propose a granular strategy for Ta-Te nanowires that successfully realizes superconductivity under high pressure and achieves a record-high upper critical field. Superconductivity emerges at 10.6 GPa, with a $T_c$ of about 3.5 K. $T_c$ exhibits a slight dome-like behavior during compression. Ta-Te nanowire is a possible candidate for applications in high magnetic field, and provides an ideal platform for investigating the exotic mechanisms of large upper critical fields.

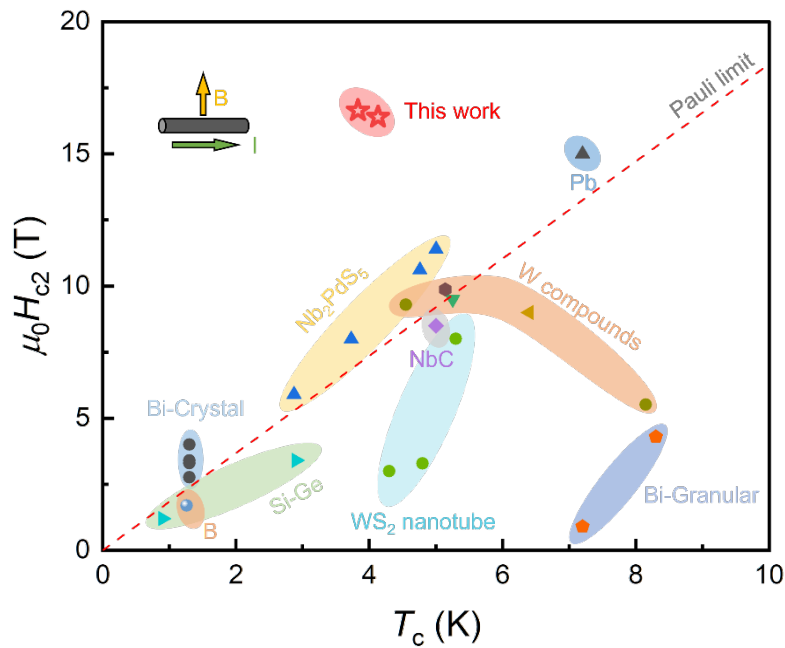

Figure 4. The upper critical fields of several superconducting nanowires. The multiple data points in one type of nanowire represent different diameters. The details are summarized in Table S1.

# Supplementary Information

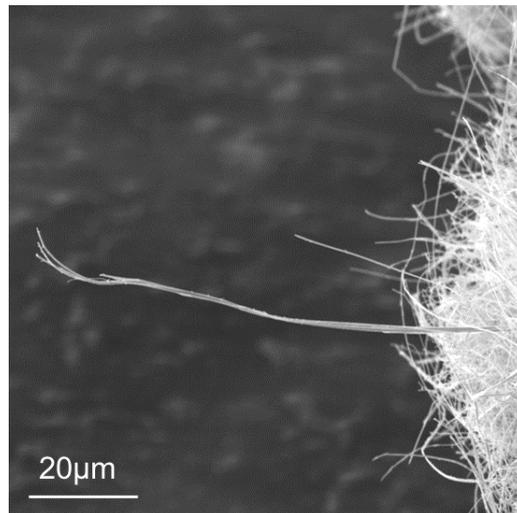

Figure S1. The SEM image of the separated nanowire for ambient resistivity measurements from a bunch of nanowires.

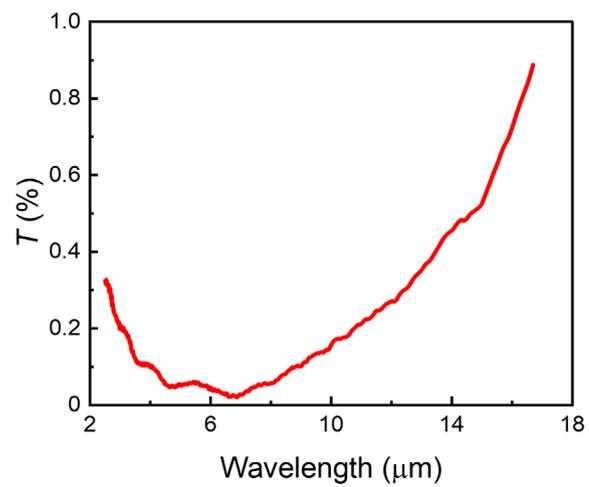

Figure S2. IR transmission spectrum of Ta-Te nanowire.

High-pressure *in-situ* electrical transport property is measured in Physical Property Measurement System (PPMS-9T) using a nonmagnetic diamond anvil cell (DAC). A cubic BN/epoxy mixture is used as insulating layer between BeCu gaskets and electrodes. The nanowires are directly collected from synthesis and then loaded into the diamond anvil cell (DAC) with 200 μm anvil culet. Four Pt foils are arranged in a van der Pauw four-probe configuration to contact the sample in the chamber for resistivity measurements[30]. Pressure is determined by the ruby luminescence method[31].

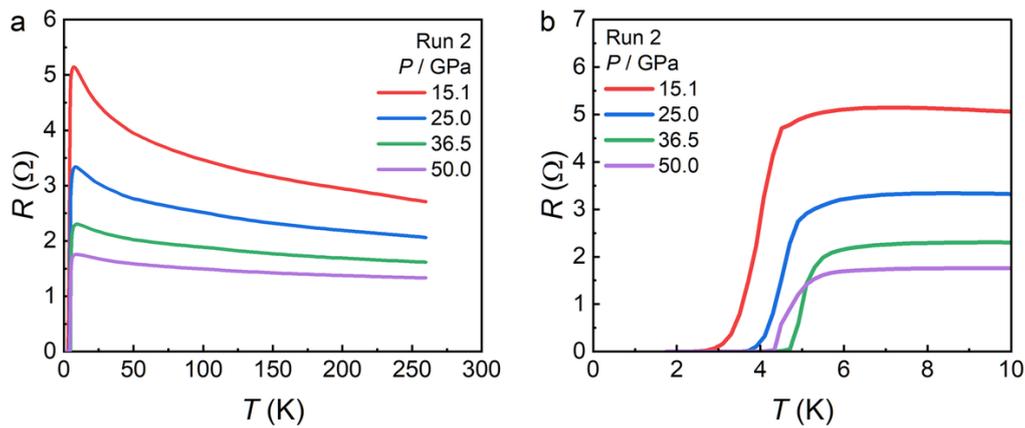

Figure S3. The temperature dependence of resistivity of Ta-Te nanowires under high pressure from Run 2. (a) 1.8 – 265 K. (b) 1.8 – 10 K.

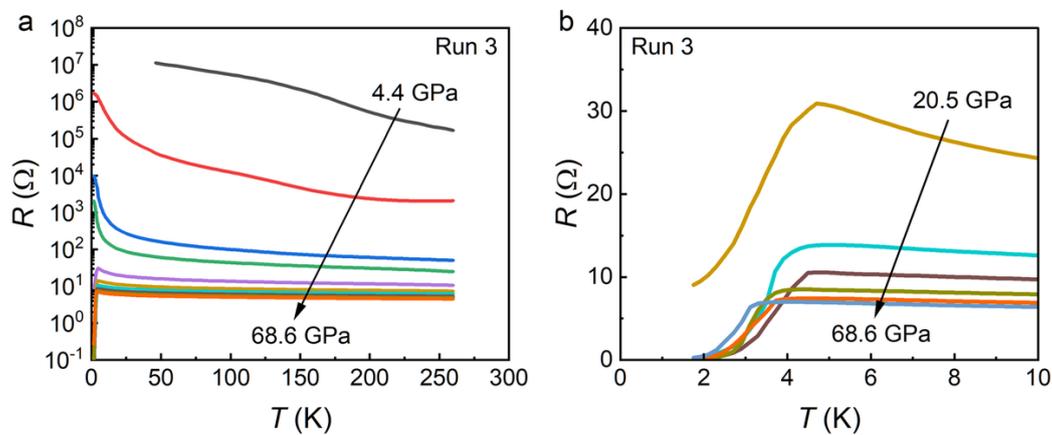

Figure S4. The temperature dependence of resistivity of Ta-Te nanowires under high pressure from Run 3. (a) 1.8 – 265 K. (b) 1.8 – 10 K.

Table S1. The upper critical field in several superconducting nanowires.

| Compound | $T_c$ (K) | $\mu_0 H_{c2}(0)$ (T) | $\mu_0 H_{c2}(0) / T_c$ | ref |
| --- | --- | --- | --- | --- |
| Ta-Te nanowires | 3.83 | 16.64 | 4.34 | This work |
| Bi | 1.3 | 4 | 3.07 | 10 |
| Ta$_2$PdS$_5$ | 5 | 11.4 | 2.28 | 32 |
| Pb | 7.2 | 15 | 2.08 | 33 |
| W(CO)$_6$ | 5.25 | 9.5 | 1.81 | 34 |
| NbC | 5 | 8.5 | 1.7 | 35 |
| W | 4.2 | 7 | 1.67 | 13 |
| He$^+$ Focused WC$_{1-x}$ | 6.4 | 9 | 1.41 | 14 |
| B | 1.257 | 1.7 | 1.35 | 12 |
| diffused Ge–Si | 2.9 | 3.4 | 1.17 | 36 |
| Bi-V | 8.3 | 4.3 | 0.52 | 9 |
| Bi-III | 7.2 | 1.0 | 0.14 | 9 |

**ACKNOWLEDGMENT**

This work was supported by the National Key R&D Program of China (Grant No. 2023YFA1607400) and the National Natural Science Foundation of China (Grant No. 52272265), The authors thank the Analytical Instrumentation Center (# SPST-AIC10112914), SPST, ShanghaiTech University.